\documentclass[twocolumn,epjc3]{svjour3}  
\smartqed  
\RequirePackage{graphicx}
\RequirePackage[numbers,sort&compress]{natbib}
\RequirePackage[colorlinks,citecolor=blue,urlcolor=blue,linkcolor=blue]{hyperref}

\journalname{Eur. Phys. J. E}
\begin{document}

\title{Measuring the average cell size and width of its distribution in cellular tissues using Fourier Transform}


\author{Tess Homan\thanksref{addr1}
        \and
        Sylvain Monnier\thanksref{addr1} 
         \and
        C\'ecile Jebane\thanksref{addr1}
         \and
        Alice Nicolas\thanksref{addr2}
         \and
        H\'el\`ene Delano\"e-Ayari\thanksref{e1,addr1}
}

\thankstext{e1}{e-mail: helene.ayari@univ-lyon1.fr}


\institute{Biophysique, ILM, Universit\'e Claude Bernard Lyon 1, Villeurbanne, France \label{addr1}
           \and
            Univ. Grenoble Alpes, CNRS, CEA/LETI-Minatec, Grenoble INP, LTM, F-38054 Grenoble-France \label{addr2}
}

\date{Received: date / Accepted: date}

\abstractdc{
 We present an in-depth investigation of a fully automated Fourier-based analysis to determine the cell size and the width of its distribution in  3D biological tissues. The results are thoroughly tested using generated images, and we offer valuable criteria for image acquisition settings to optimize accuracy.  We demonstrate that the most important parameter is the number of cells in the field of view, and we  show that accurate measurements can already be made on volume only containing 3x3x3 cells. The resolution in $z$ is also not so important and a reduced number of in-depth images, of order of one per cell, already provides a measure of the mean cell size with less than 5\% error. The technique thus appears to be a very promising tool for very fast live local volume cell measurement in  3D tissues \textit{in vivo} while strongly limiting photobleaching and phototoxicity issues. 
}

\maketitle

%
%
\section{Introduction}\label{sec:intro}
Volume is a key parameter in various fundamental biological processes such as cell growth, division or fate \cite{Zlotek-Zlotkiewicz2015,Guo2017}. It is tightly regulated during cell cycle \cite{Zlotek-Zlotkiewicz2015} and is dependent on the cellular microenvironnement chemical and physical properties \cite{Hoffmann2009, Guo2017}. It can be regulated by cellular tension \cite{PerezGonzalez2018}. Different techniques have been used for measuring isolated cells volume such as 3D cell reconstruction \cite{Guo2017}, Fluorescence Exclusion measurements \cite{Cadart2017} and commercial coulter counter \cite{Bryan2012}. The regulation of cellular volume within 2D/3D tissues is unknown so far, mainly due to technical limitations  associated with its measurement and follow up in time.

Indeed, obtaining cell area/volume necessitates precise cell segmentation of its boundaries in 3D, which most often requires the acquisition of a z-stack of a sample in which the membrane \cite{Bosveld2012}  or the intercellular space \cite{Marmottant2009} have been made fluorescent. However, the imaging of living samples is restricted by cell movements, photobleaching or even cell death (photo-toxicity). Capturing an entire $z$-stack can take up to several minutes at high resolution, which is on the same timescale as the rearrangement of cells inside the tissue. In addition, cells are very sensitive to light, with overexposure leading to cell damage. The amount of images which can be acquired to create a $z$-stack is therefore limited by the amount of  light cells can handle. So developing a new tool which would require the minimum z slices acquisition is of real importance for time follow up of \textit{in vivo} tissues.

Based on previous work \cite{Durande2019a}, we propose here a 3D Fourier transform tool for live measurement of cell volume within 3D tissues. Fourier analysis describes the image as a superposition of sinusoidal functions (repetitive patterns with a set frequency) \cite{Bracewell1986}. Most images are a combination of many frequencies, but cellular tissues are made up of distinct units: cells. Therefore the result of the Fourier analysis has one dominant frequency corresponding to the average cell size. We tested here how accurate this method can be by using \textit{in silico} simulated data of 3D tissues. This also allowed us to independently vary all important parameters, such as the number of cells in field of view, cell size homogeneity, and acquisition resolution. Based on this analysis, we propose protocols to help define image acquisition parameters, in the form of a set of rules and easy-to-read graphs. The method is then tested on live cell aggregates, which represent good \textit{in vitro} tumor models \cite{Costa2016}.


\section{Materials and Methods}\label{sec:MM}
\subsection{Cellular aggregates preparation}
We use HT29 cells for these experiments. Cells are culture in  DMEM Medium (GIBCO 61958-026) supplemented with 10$\%$ FBS (Pan Biotech P308500) and 1$\%$ Penicylin-Streptavidin (GIBCO,15140-122). They are maintained at 37$^o$C with 5$\%$ CO$_2$ and passaged twice a week. Aggregates are formed using Ultra Low Adhesion 96 Well Plates (Greiner bio-one, 650970). After passaging, cells are counted and diluted to a concentration of 10 000 cells/mL. From this stock solution, a desired number of cell is subsequently seeded into a well (for example 300 cells/mL) and left to grow into aggregates for a minimum of 48h.

\begin{figure*}
\includegraphics[width=\textwidth]{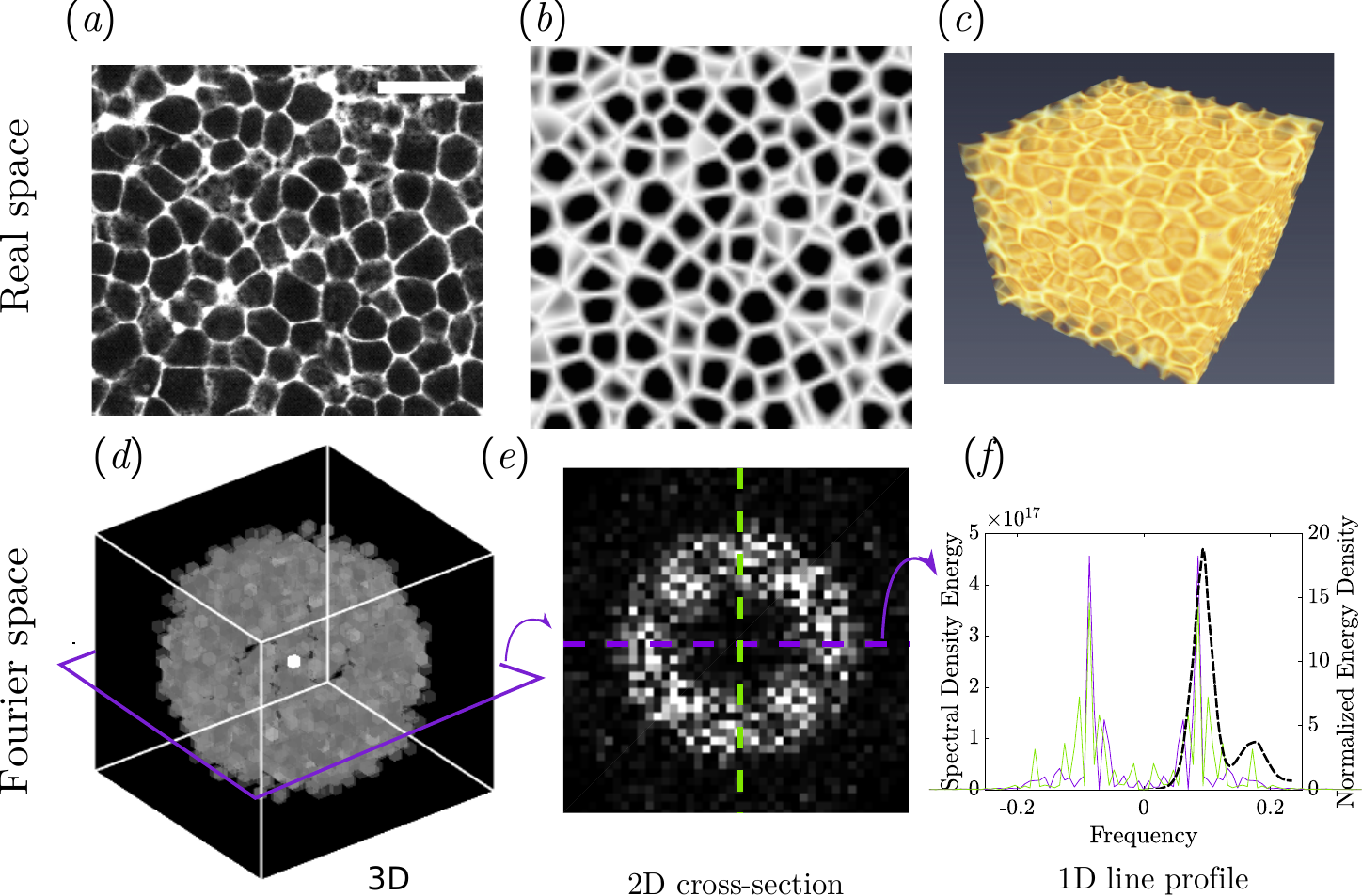}
 \caption{(a) Typical z-slice of an HT29 aggregate using  two-photon acquisition. Bar: 30 $\mu$m. (b)  Z-slice  ($128 \times 128$ pixels) of a simulated 3D aggregate generated using the Voronoi method. The cells have an average radius of 11 pixels.  Note that faces parallel to the z-slice can result in large white areas.  (c) 3D rendering of the entire z-stack where $dz = 1$  pixel i.e. the same resolution as  in $x, y$. (d) Fourier transform of (c).  A bright center pixel and a light spherical shell are visible. (e) 2D cross section of (d). The magnitude of the central peak is set to zero for rendering purposes. (f)  Spectral density energy along the purple  and green dashed line displayed in (e). The black dashed line is the normalized power spectral density $e_n(k)$ as defined in Eq. (\ref{eq:I2agregat}).  The position of the main black peak provides an accurate measurement of the mean cell size.}
\label{fig:fig1}     
\end{figure*}

\subsection{Two-photon imaging of cellular aggregates and application of osmotic shocks}
Aggregates were imaged either in simple wells with $\#$1 coverslip at the bottom or in $40 \mu m$-high microfluidic channel. Entrapment in the microfluidic channel prevented aggregates from moving around and simplified media changes. 1 h after the injection of the aggregates in the well or in the microfluidic chamber, the extracellular space was stained with FITC-Dextran diluted at  2 mg/ml thus staining the interstitial fluid within the aggregate. All experiments were conducted at 37$^o$ in CO$_2$-independent cell culture  medium (GIBCO 18045-088). Imaging was performed using a two-photon setup on a Nikon microscope equipped with a 780 nm laser. Osmotic shocks were applied by adding  6 kDa Dextran at a concentration of 100 g/L (for details see \cite{Dolega2021}).

\subsection{Artificial cell  images generation}
Synthetic 3D images of size $L\times L \times L$ pixels mimicking real tissues were generated. Random points (representing the centers of the cells) are added automatically one after the other in a 3D matrix. For each new point, a  radius is randomly taken from a normal distribution centered on $r_0$ with a width $\sigma$. A new point, $i$, is only added in the lattice if the sphere centered on $i$ with radius $r(i)$ fits in with all other spheres. If not, a new random point is chosen. When the number of spheres increases, the probability that the new sphere fits in the free space decreases. We thus used as a cutoff a maximum number of attempts to add a cell. Note that as the space has a very low probability of being fully filled with compact spheres, the obtained lattice is not regular even when $\sigma$ is set to $0$. The width of the size distribution was adjusted in two ways, either by changing $\sigma$  or by changing the number of possible attempts for filling the lattice. As $r_0$ defines the typical number of pixels per cell, changing this value enabled to  simulate different magnifications used when acquiring real data with different objectives at different resolutions. The total number of cells that are visible in the image were set by changing $L$, which corresponds to changing the field of view in data acquisition. To avoid edge effects, the actual space in which the centers were distributed runs from $-2r_{max}$ to $L+2r_{max}$ in $x, y, $ and $z$-direction.
The images were then cut out from the center to create an $L\times L \times L$ $z$-stack.

 Cell boundaries were then generated. More precisely, a 3D Voronoi tesselation was performed, using cell centers as seeds.  As experimentally the cells edges or faces are larger than 1 pixel (Fig. \ref{fig:fig1}a), the grey values of the pixels in every Voronoi cell were scaled with the relative distance to the seed. This allowed obtaining a 3D image much closer to real data (Fig. \ref{fig:fig1}b). Amira software was used to segment and visualize the simulated data (Fig. \ref{fig:fig1}c).

\subsection{Fourier Transform Analysis}
Fourier transformation was used to analyze the intensity of the 3D images. The Fourier transform of the intensity is expected to resemble the diffracted intensity in an isotropic medium scatterred by elastic processes. Before applying Fourier Transform, the images were decomposed in periodic ($p$) and smooth components ($s$). The Fourier transform was applied on  $p$ only to avoid edge effects artefacts \cite{Moisan2011a}. The discrete Fourier Transform (DFT) was computed using a Fast Fourier Transform (FFT) algorithm \cite{Frigo1998} with convention:
\begin{equation}
\tilde{I}(\vec{k}) = \sum_{l,m,n=0}^{L-1} I(\vec{x}) \exp(- 2 i \pi \vec{k}\cdot \vec{x})
\end{equation}
with $I$ the intensity, $\vec{x}$ the discrete position indexed by $(l,n,m)$ and  $\vec{k}=1/L*(s,t,u)$ the discrete wave vector ($L$ being dimensionless). 

Consistent with elastic diffusion processes of slightly ordered isotropic materials, a bright shell around an even brighter center pixel was obtained (Fig. \ref{fig:fig1}d). We expect that the radius of the shell , $k_D$, can be used to obtain an estimation of the averaged cell diameter. To properly determine $k_{D}$, we took advantage of the spherical symmetry of the FFT. $\tilde{I}$ was interpolated in spherical coordinates ($k$,$\theta$,$\phi$). The normalized power spectral density of the cell aggregate is calculated:
\begin{equation}\label{eq:I2agregat}
e_n(k)= \frac{k^2 \int |\tilde{I}(\vec{k})|^2\sin \theta\, d\theta\,d\phi}{\int |\tilde{I}(\vec{k})|^2 d\vec{k}}
\end{equation}
Different from conventions used in the analysis of elastic diffusion of light, Equation (\ref{eq:I2agregat}) is normalized by the total power spectral density and not by the spectral volume. $e_n(k)$ then shows well emerged peaks that we considered to determine the average cell size in the tissue (Fig. \ref{fig:fig1}f).

\subsection{Calculation of cell form and structure factors}
The scattering pattern should be linked to a cell form factor that informs on the mean symmetry properties of the cells and a structure factor that informs on the cellular arrangement. We calculated both factors independently on simulated images. Cells that were entire in the image volume (not cut by the borders) were considered. The centroid of the cells (which could differ from the points used to generate the Voronoi tesselation) were extracted and masks were generated to display each cell individually (Fig. \ref{fig:struct_fact}). The power spectral density was calculated for each single cell as in Eq. (\ref{eq:I2agregat}). The cell-averaged power spectral density of the single cells was then obtained:
\begin{equation}\label{eq:I2cell}
e_{n,cell}(k)= N_{cell}\frac{  < k^2 \int_{cell_i}|\tilde{I}(\vec{k})|^2 \sin \theta\, d\theta\,d\phi>_{cells} }{\int |\tilde{I}(\vec{k})|^2 d\vec{k}}
\end{equation}
with $N_{cell}$ the number of entire cells in the image.

\section{Results}\label{sec:Res}

\subsection{Cell sizes}\label{sec:size}
Figure \ref{fig:fig1}d-e shows an isotropic diffusion pattern. This suggests that the cells in the aggregate are only weakly ordered, the distance of the shell from the center providing the mean distance in between the cells centers. This distance corresponds to the cell diameter for 3D compact aggregates as we address here. As the relationship between the cell distance and the frequency peak is very sensitive to the accuracy of the model used to describe the cell aggregate \cite{Kinning1984}, we first wrote $D_F = \gamma /k_{max}$ with $\gamma$ a factor of proportionality.  The cell size given by the peak of the normalized Fourier energy density and the mean cell size directly obtained in real space on the artificial images were then compared. However, even though for the original data the exact locations of all cell walls are known, it is still not trivial to define an averaged cell size. From the Voronoi tesselation we easily get the volume of each cell (completely enclosed in the image), whose averaged value is denoted $V$.  The averaged diameter was thus obtained from dimensional analysis: $D_{V}=\lambda V^{1/3}$, the proportionality factor $\lambda$ depending on the precise mean shape of the cells. Choosing a cell shape between spherical and cubical would alter the calculation of the diameter by 30\%. As in the cell volume calculation the contour is always included, we will more precisely use $D_{V}=\lambda V^{1/3}-1$ (pixels) so that the cell contour is not counted twice in the distance between two cells.

\begin{figure}
  \centering
  \includegraphics[width=0.8\linewidth]{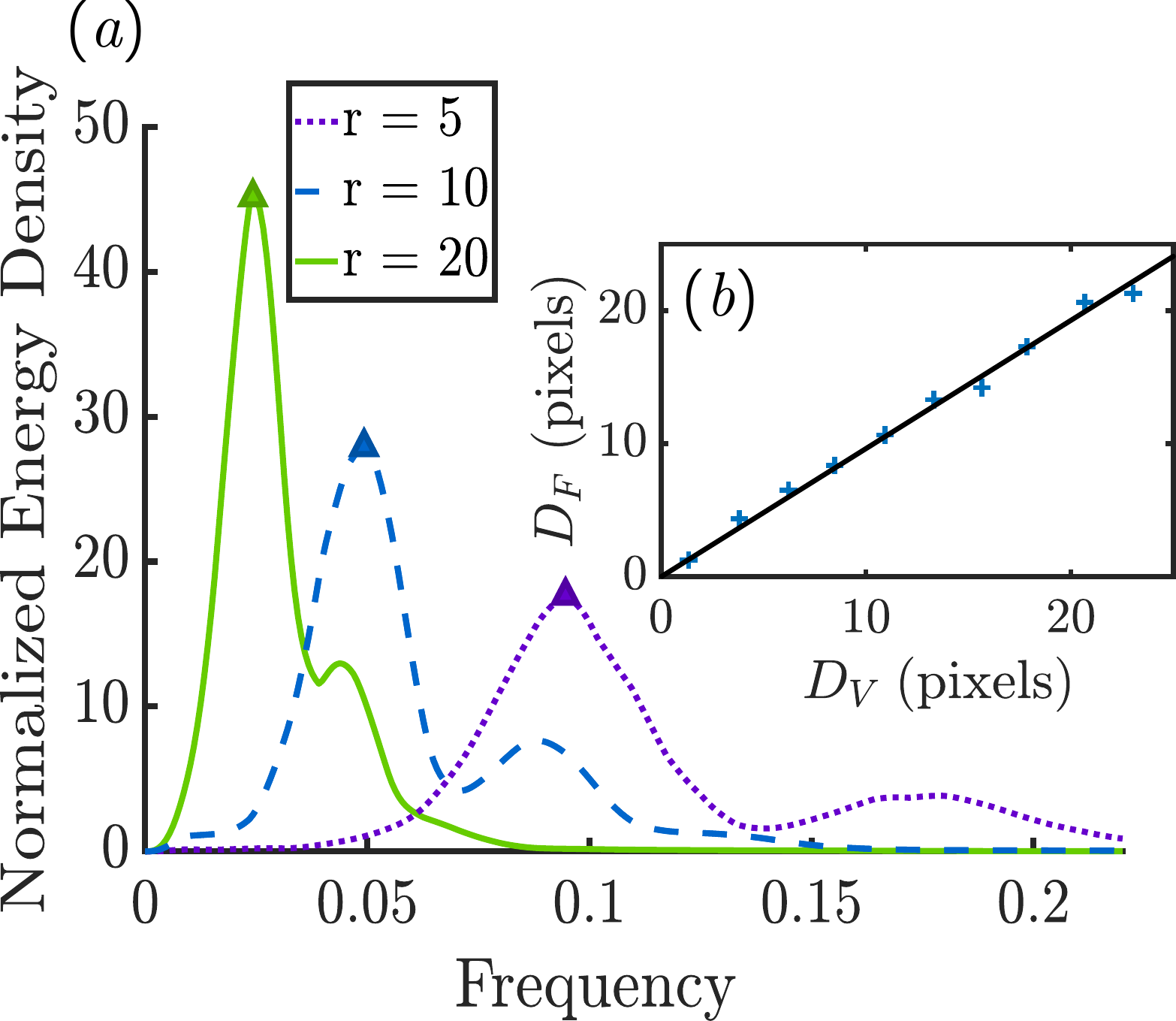}%
  \caption{$(a)$ Curves resulting from the Fourier analysis for three different values of $r_0$, $\sigma =0$, $L=128$. Position of the peaks corresponds to the frequency or wavelength of the artificial tissue. $(b)$ Plot of  $D_F$ as a function of the original $D_{V}$, with an affine fit with a slope of 0.96 ($r^2=0.992$). }
    \label{fig:r}
\end{figure} 

Several values of $r_0$, the mean radius  of the Voronoi cells, were tested (Fig. \ref{fig:r}).The peak was both wider and lower when increasing cell size, due to finite size effects in the Fourier transformation. Figure \ref{fig:r}b shows that the measure from the FFT analysis correlates very well with the direct measure from the Voronoi cells  with choosing $\gamma = 1$ and $\lambda=1$ (slope $0.96$, coefficient of determination $0.992$).

\subsection{Relation between the energy spectral density, the structure factor and the cell form factor\label{sec:structfact}}

\begin{figure}
  \includegraphics[width=1\linewidth]{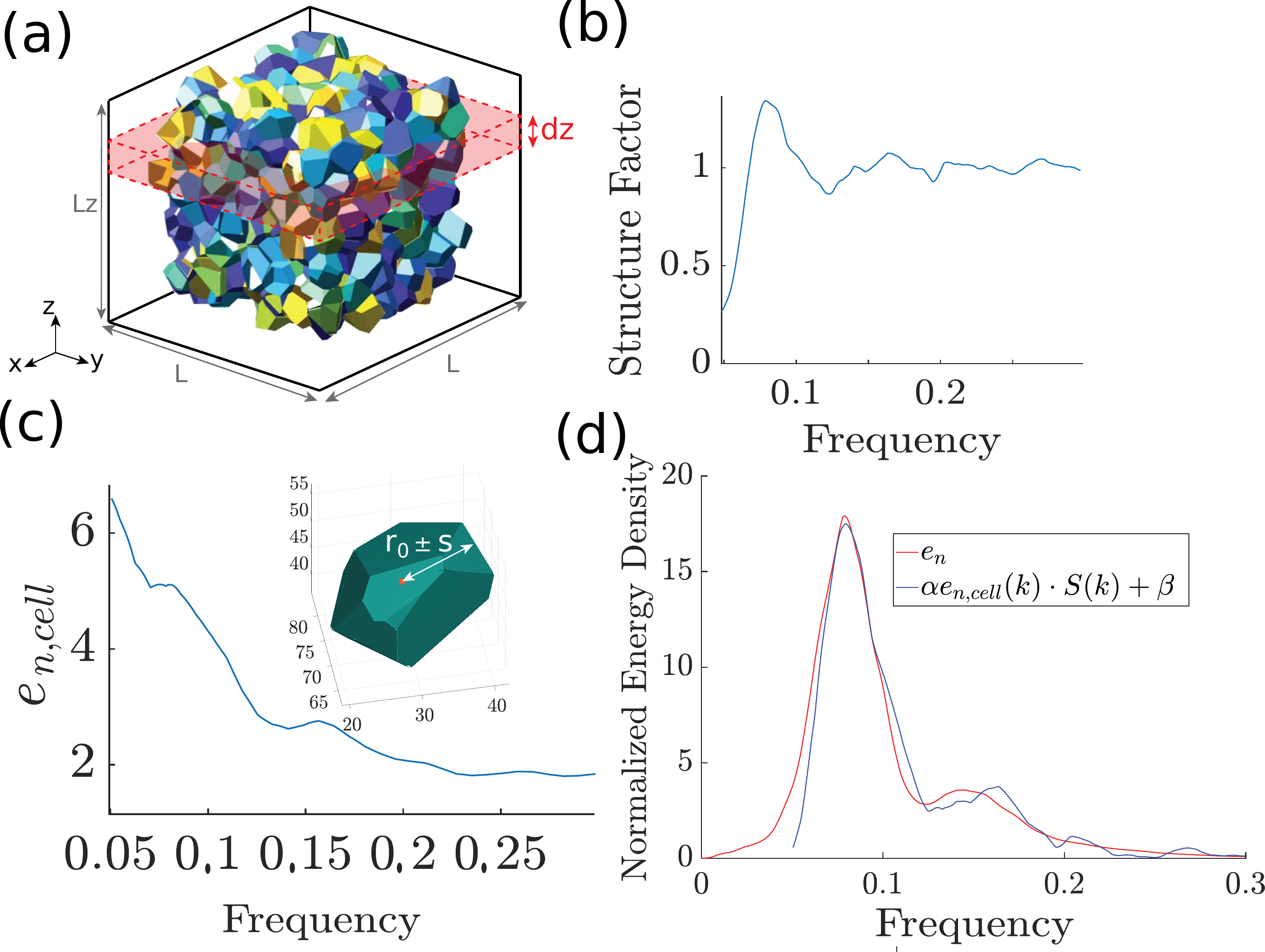}%
  \caption{ $(a)$ From the Voronoi tesselation each cell in the image volume could be individually segmented to obtain both its centroid and its associated cell form factor. $(b)$ Structure factor computed from the centroid obtained from $(a)$ .$(c)$ Cell form factor averaged from all the cells segmented in $(a)$.$(d)$ Comparison of the energy density of the cell image and the average cell form factor multiplied by the structure factor, both quantities show a linear relationship with a coefficient $\alpha=3.45$ and $\beta=-6.15$.
    }
    \label{fig:struct_fact}
    \end{figure}
    
 Form and structure factors are generally deduced from diffusion patterns and help characterizing the organization of the matter. We tested here whether the spectral energy densities introduced in Equations (\ref{eq:I2agregat}) and (\ref{eq:I2cell}) allow to deduce a value for these two factors. The power spectral density is now calculated for the $N_{cell}$ cells of the 3D aggregate within the approximation of  isotropic material. We denote $f_j(\vec{r})$ the fluorescence intensity due to cell $j$. Within the approximation that all the cells have the same intensity profile $f(\vec{r})$, the total intensity is:
\begin{equation}
I(\vec{r})=\sum_{j=1}^{N_{cell}} f(\vec{r}-\vec{r_j})
\end{equation} 
with $\vec{r_j}$ the center of cell $j$. The square norm of $\tilde{I}(\vec{q})$ then writes:
 \begin{equation}
    |\tilde{I}(\vec{k})|^2=N_{cell}*|\tilde{f}(\vec{k})|^2S(\vec{k})
  \label{eq:formstruct}
 \end{equation}
 with $S(\vec{k}) = \frac{1}{N_{cell}}\sum_{j=1}^{N_{cell}}e^{-2\pi i\vec{k}.(\vec{r}_j-\vec{r}_k)}$ the structure factor and $|\tilde{f}(\vec{k})|$ the cell form factor.
Averaging over all angles, and if no correlation exists in between $S(\vec{k})$ and $|\tilde{f}(\vec{k})|$,  the power spectral density thus writes:
\begin{equation}\label{eq:enecell}
e_n= e_{n,cell}(k) \cdot S(k)
\end{equation}
with 
\begin{equation}
e_{n,cell} =\frac{ 4 \pi N_{cell} k^2 |\tilde{f}(k)|^2}{\int |\tilde{I}(\vec{k}) d\vec{k}}
\end{equation} 
being the averaged power density of all individual cells renormalized by the total spectral density of the image and $S(k)= \int S(\vec{k})\,sin \theta\, d\theta\,d\phi$  the integrated structure factor on a shell of radius $k$. Figure \ref{fig:struct_fact}b shows the integrated structure factor calculated for the cells in the simulated data. The latter shares similarities with the structure factor of a hard sphere fluid, but still appears more complex. This suggests that the interaction potential between the cells requires refinement compared to volume exclusion. Unfortunately, both the cell form factor (Fig. \ref{fig:struct_fact}c), and the structure factor (Fig. \ref{fig:struct_fact}b) cannot be obtained analytically as it could have been the case for simpler geometries and known interactions (see \cite{Pedersen1997} for a review of well known cases where models do exist).

The power spectral density was fitted using Equation (\ref{eq:enecell}). This could only be done by introducing two constants, $\alpha$ and $\beta$, that evidenced an affine relationship between $e_n$ and $e_{n,cell}$:
\begin{equation}\label{eq:en_fit}
e_n(k)=\alpha e_{n,cell}(k)\cdot S(k)+\beta
\end{equation}

We obtained $\alpha=3.45$ and $\beta=-6.15$ (Fig. \ref{fig:struct_fact}d). The fact that we do not get $\alpha=1$ and $\beta=0$ may  arise from the conjunction of different reasons.  Firstly, cross-correlations between the shape and the structure factors are expected, as cells are deformable objects and are packed in the aggregate. Secondly,  the cell shape factor is accounting twice for each cell facet (see material and Methods). Lastly, but maybe the most important reason, this adjustment may originate from finite size effects, the numerical Fourier Transform being a truncated version of the analytical one. However Figure \ref{fig:struct_fact}d still shows that the Fourier energy density is well rationalized using the product of the average cell form factor and the structure factor. Thus it clearly appears that the peak in $e_n$ corresponds to the first peak of the structure factor. 
    
\subsection{Effect of experimental parameters on cell size measurements}
Below the robustness of the Fourier analysis of the energy density to provide an accurate measurement of the mean cell size is tested. Cell number and image resolution are varied, to mimic variations of the magnification and the resolution of the imaging. For all the results that follow, a set of 11 3D images generated with $r_0=5$ (i.e. an average cell size $D_{V}=10$ pixels) were used.

	\subsubsection{Effect of the number of cells per field of view}\label{subsection:imsize}

\begin{figure}
  \centering
  \includegraphics[width=\linewidth]{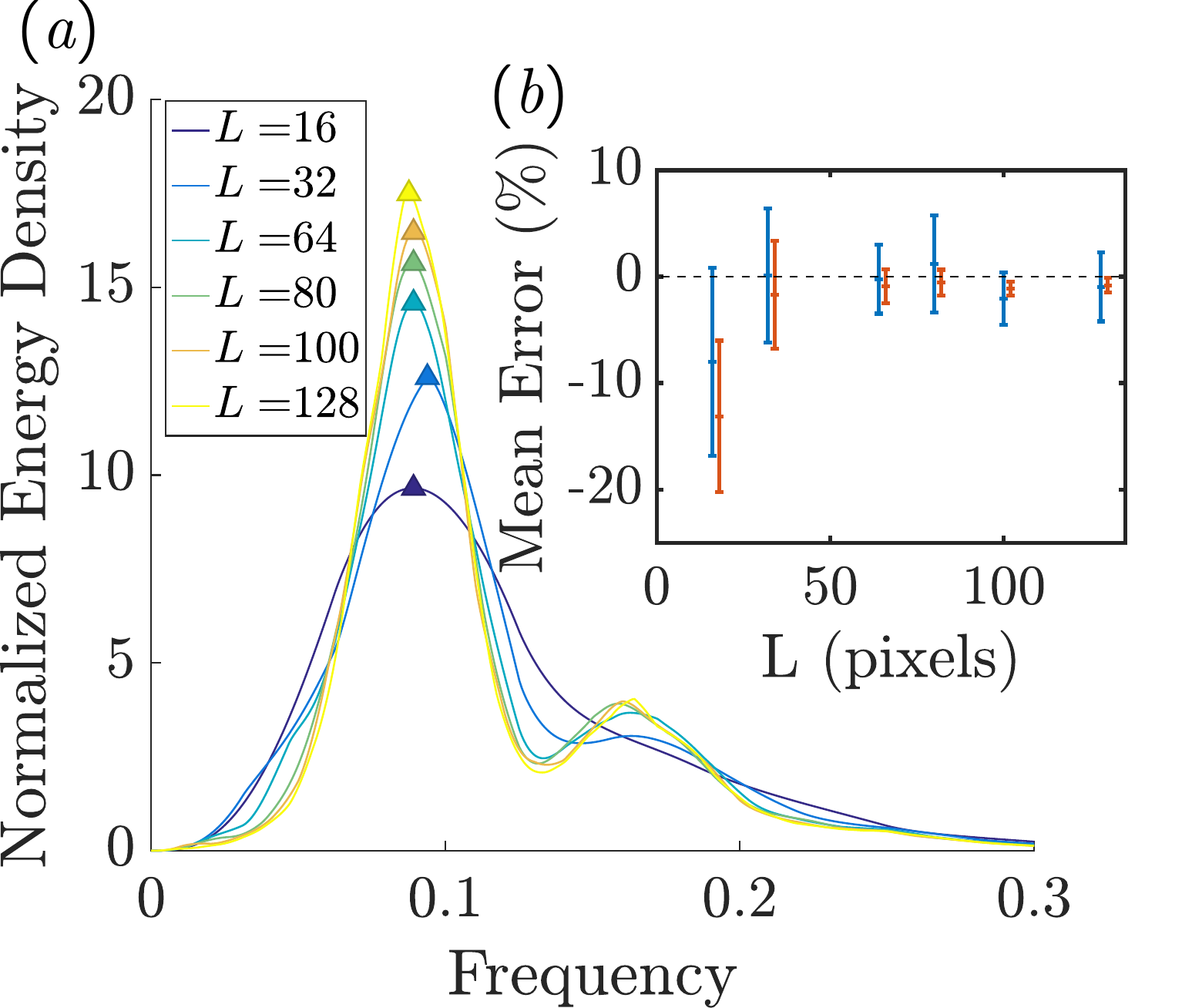}%
  \caption{ $(a)$ Shape of the normalized energy density as a function of the size $L$. As expected, the peaks become smaller and wider when $L$ decreases.  $(b)$ Blue: Mean error $(D_F-D_{V})/D_V$ in $\%$ and its associated standard deviation calculated on a batch of 11 3D images  as a function of the size $L$ ($r_0=5$, $\sigma$ between 0 and 2). Red: Same for $(D'_F-D_{V})/D_V$. }\label{fig:Lerror}
\end{figure}

From the point of view of the Fourier transform, the most important parameter is the number of cells in the field of view, in both $x$, $y$, and $z$, far above the number of pixel per cell, as we already demonstrated in 2D \cite{Durande2019a}.
The higher the number of visible wavelengths, the stronger the peak at that corresponding frequency in Fourier space (see Fig. \ref{fig:Lerror}b).  Influence of the cell number and more specifically quantification of the minimal number of cells that are required for the effectiveness of the FFT method were investigated. 

The mean error on cell size measurement was investigated as a function of the size $L$ of the image (taken the same in $x$, $y$ and $z$) (see Fig.\ref{fig:Lerror}b). Even when $L=32$ the mean error remains close to zero with standard deviation  around $3\%$. As cell size is set around 10 pixels, this means that 3 cells in each direction in the image were enough to obtain a valid  result.This leads to the conclusion that this method is efficient for providing the mean cell size even with high magnification images, where the number of cells is reduced.

As the direct measure of the position of the peak in Figure \ref{fig:fig1}f is sensitive to the noisy shape of the curve and may lead to an imprecise determination of the cell size, the position of the peak was also calculated as the center of the full width at half maximum:
\begin{equation}\label{eq:D'F}
f_{peak}=\frac{\int_{f_1}^{f_2}k*e_n(k)\,dk}{\int_{f_1}^{f_2}e_n(k)\,dk}
\end{equation}
with $f_1$ and $f_2$ the frequencies  at half height of the principal peak. Equation (\ref{eq:D'F}) thus leads to a mean cell diameter $D'_{F} = 1/f_{peak}$. Figure \ref{fig:Lerror}b shows that the cell diameter calculated by this latter technique provides slightly underestimated values compared to the direct measure of the peak position, but however with a reduced dispersion showing that it is more robust.

\subsubsection{Effect of the pixel size in $z$}\label{sec:dz}

\begin{figure}
  \centering
  \includegraphics[width=0.8\linewidth]{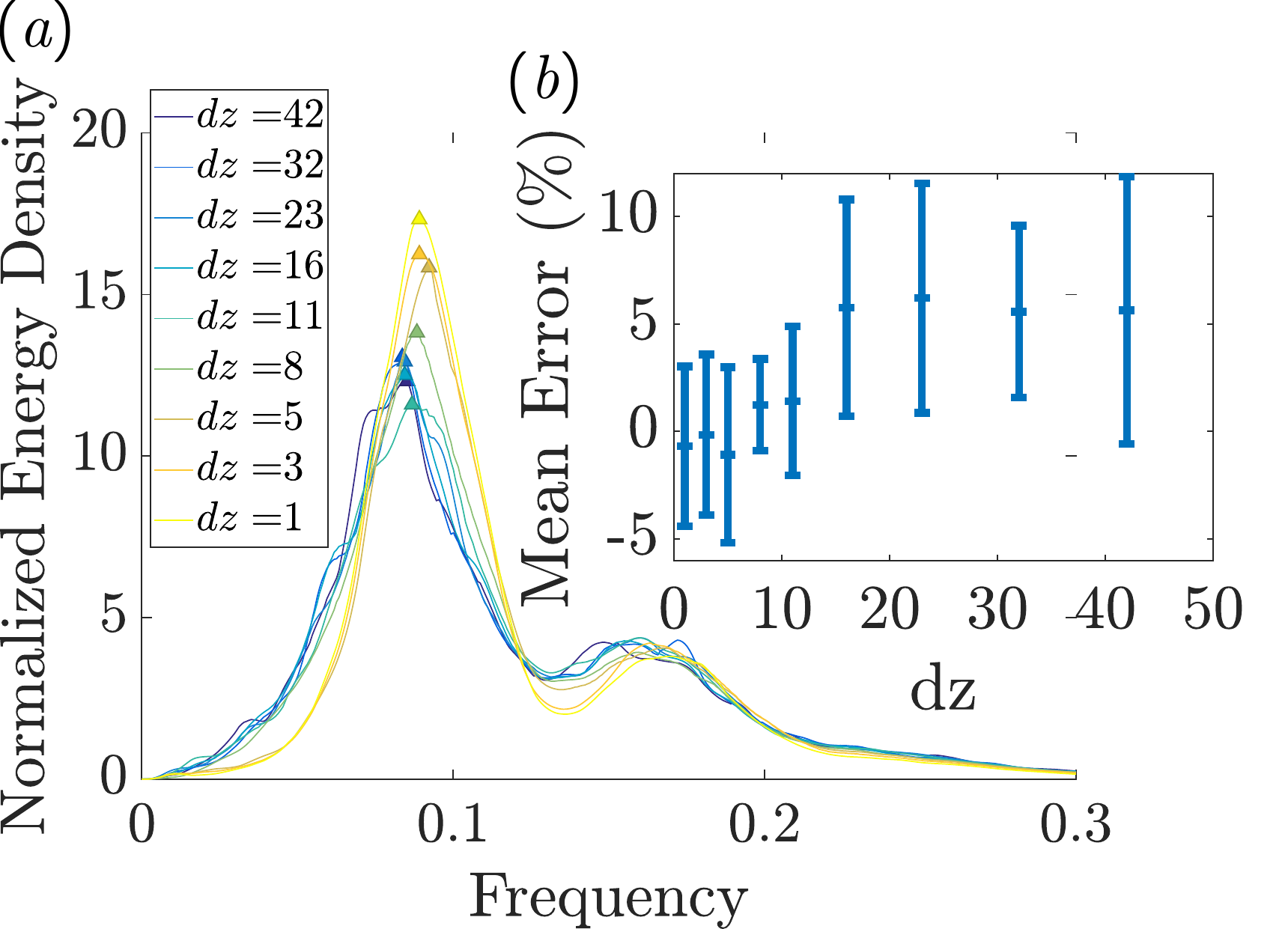}%
  \caption{$(a)$ Shape of the normalized energy density as a function of frequency for in-depth pixel size $dz$. $(b)$ Mean error $(D_f-D_{V})/D_V$ in $\%$ and its associated standard deviation calculated on a batch of 11 3D images as a function of the in-depth pixel size $dz$ ($r_0=5$, $\sigma$ between 0 and 2, $L=128$).}\label{fig:zdz}
\end{figure}
From an image acquisition perspective, one of the hardest things to obtain is a high resolution in $z$-axis direction. So far the resolutions in $x$, $y$, and $z$ we used were identical. Experimentally, the distance between in-depth layers, $dz$, is twice, or even four times the resolution in $x$ and $y$. Influence of the in-depth resolution in the measurements was tested by removing information from a simulated 3D image with  $128 \times 128 \times 128$ pixels. Only one slice every $dz$ was kept, $dz$ representing the pixel size in $z$. As shown in Figure \ref{fig:zdz}b, the influence of $dz$ becomes significant when $dz$ is of order of the cell size. The results are then biased, and cell sizes are overestimated, with a larger error. This result suggests that this method keeps accurate even with standard microscopy tools and does not requires high resolution microscopy.

\subsubsection{Effect of sample thickness}

\begin{figure}
  \centering
  \includegraphics[width=\linewidth]{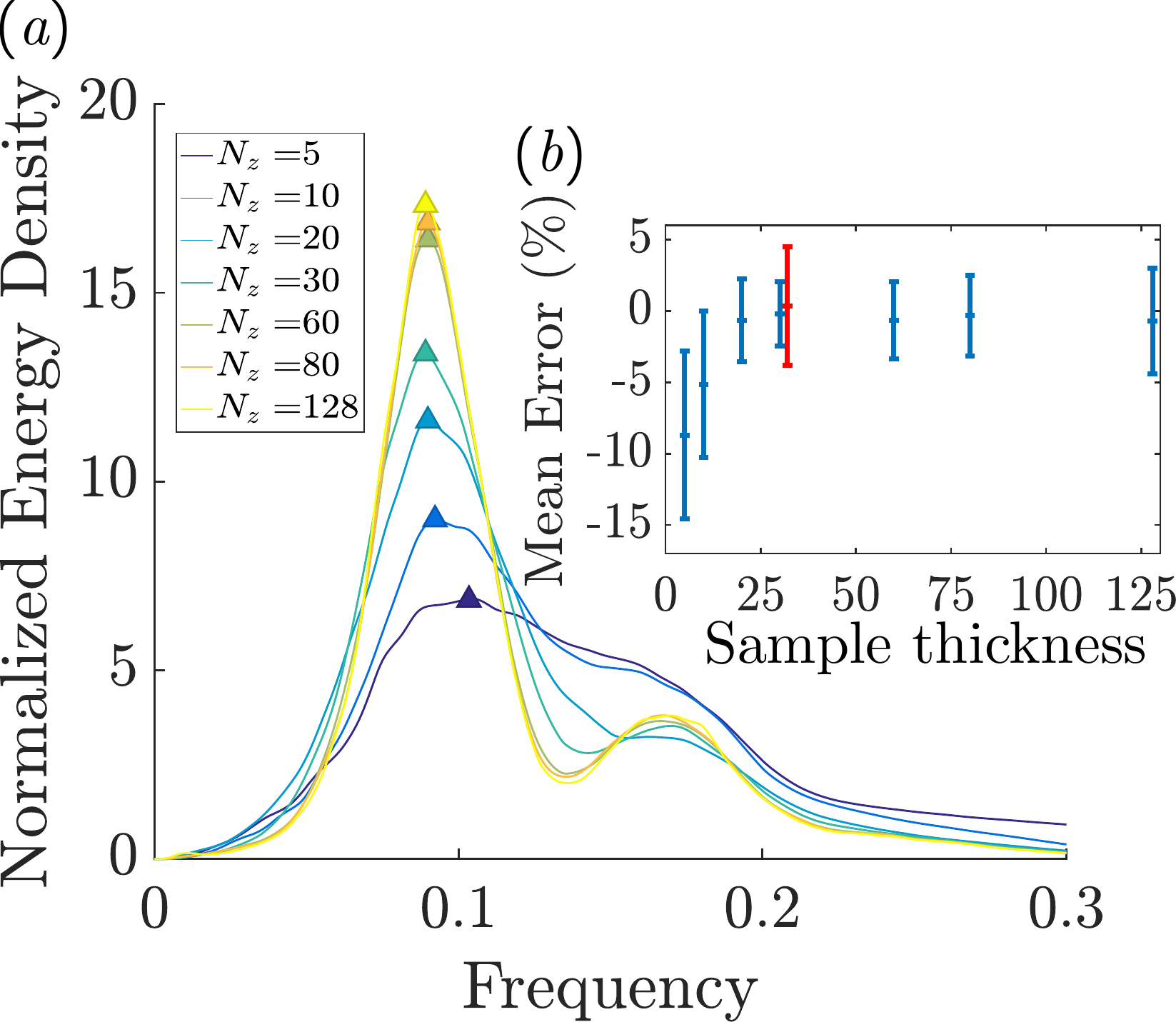}%
  \caption{$(a)$ Shape of the normalized energy density as a function of frequency for different numbers of $z$ slices. As expected, the peaks becomes smaller and wider while this number decreases. $(b)$ Mean error  and its associated standard deviation calculated on a batch of 11 3D images as a function of the number of $z$ slices in the stack ($r_0=5$, $\sigma$ between 0 and 2, $L=128$). Blue points, $dz=1$. Red point: $dz=10$. }\label{fig:ztoterror}
\end{figure}

Experimentally, the total number of $z$ layers in the stack is often significantly lower than the total number of pixels in $x$ and $y$ directions. We wondered if 3D is really necessary or if a single cross section could be enough to get an accurate measure of the mean cell size. Figure \ref{fig:ztoterror} shows that as the number of $z$ slices decreases the results are more and more biased. The cells appear smaller than they really are (which is consistent as within a single plane, many sliced cells are visible, that appear smaller than they really are). Consistent with Section \ref{subsection:imsize}, the optimal number of slices to get an accurate measurement  is about 30, which corresponds to about three cells. Altogether, this suggests that stacks that show only 3 cells in all directions and have an in-depth resolution of order of the cell size (thus a $z$-stack of 4 images) is enough to get a proper measurement of the cell size. Figure \ref{fig:ztoterror}b shows the accuracy of the measure in these conditions (red point). The mean error is close to zero although the standard deviation is slightly larger than for a more resolved imaging.

\section{Measuring the standard deviation of the size distribution}\label{sec:std}

\begin{figure}
  \centering
  \includegraphics[width=1\linewidth]{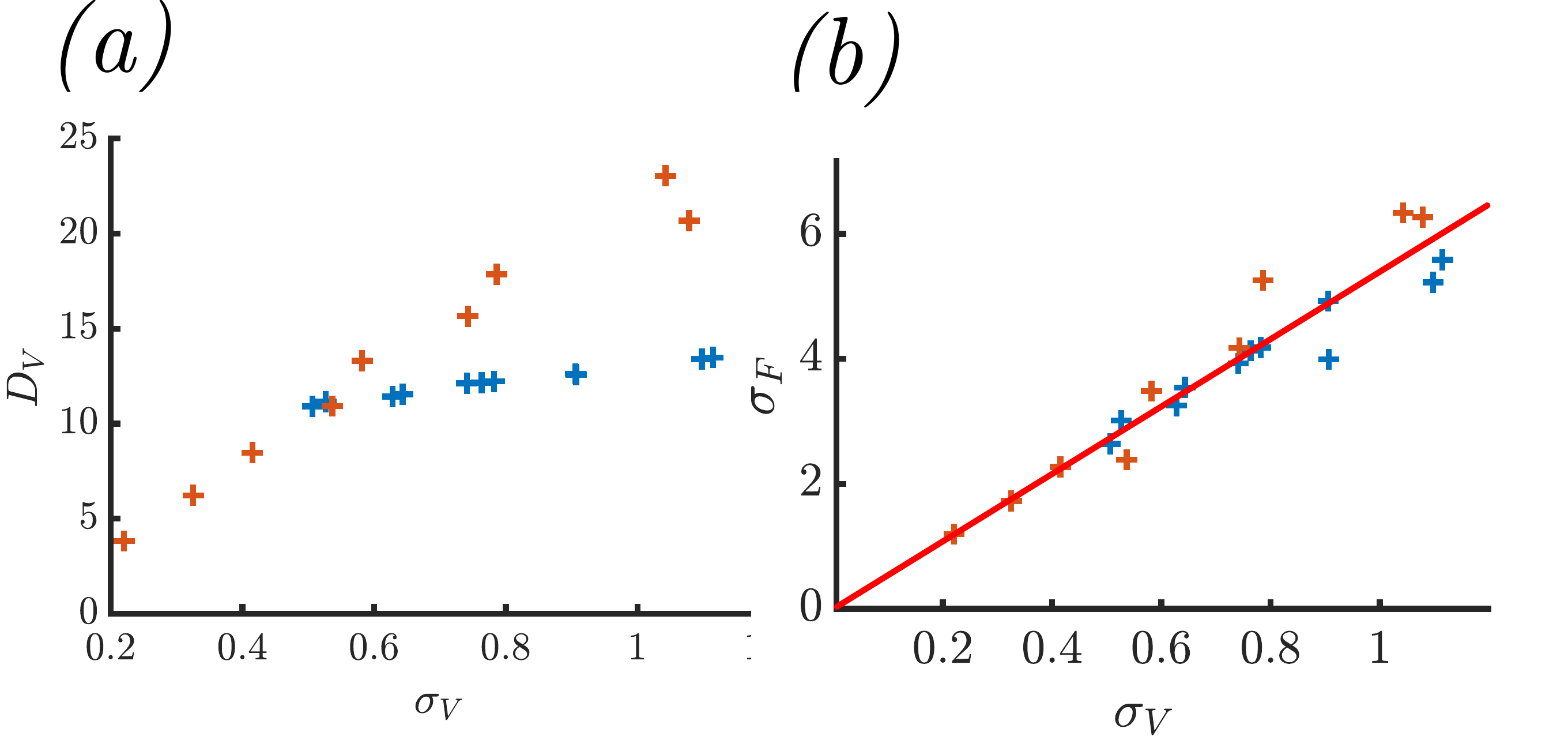}%
  \caption{$(a)$ Width of the size distribution as a function of the average size for two different sets of artificial images (in red and in blue). $(b)$ Proxy for the width of the distribution obtained from the Fourier analysis (Eq. (\ref{eq:sF})) as a function of the actual width obtained from segmented images. Both correlate linearly, with a slope $a=5.4$ (red line). Determination coefficient of 0.9. } \label{fig:width}
\end{figure}

The quantification of the standard deviation of the cell size distribution was addressed empirically. Two different batches of simulated data were considered (see Materials and Methods) as for a given set the width of the distribution is proportional to the mean size (Fig. \ref{fig:width}a) which may lead to undesired correlations. In the first set of data, the distribution of the distance between the Voronoi seeds was a Dirac function, whose center was varied. In the second set, the distribution was a gaussian with a given mean distance and the width of the distribution was varied.  We show  that for both ways of generating the Voronoi cells, a good proxy for the quantification of the standard deviation of the cell size distribution is:
\begin{equation}\label{eq:sF}
\sigma_F=1/f_1-D'_F
\end{equation}
 where  $f_1$  is the frequency at half height on the left of the principal peak, $f_1<f_{D'_F}$. $\sigma_F$ was compared to the manual determination of the standard deviation of the size distribution $\sigma_V$  (Figure \ref{fig:width}b). A linear correlation between $\sigma_{V}$ and $\sigma_F$ was obtained with a high determination coefficient. This validates the use of Eq. (\ref{eq:sF}) to quantify cell size dispersion as $\sigma_F/a$ ($a=5.4$ being the correlation coefficient, see Fig. \ref{fig:width}).

\section{Results on two-photon stacks of cellular aggregates}

\begin{figure}
  \centering
  \includegraphics[width=1\linewidth]{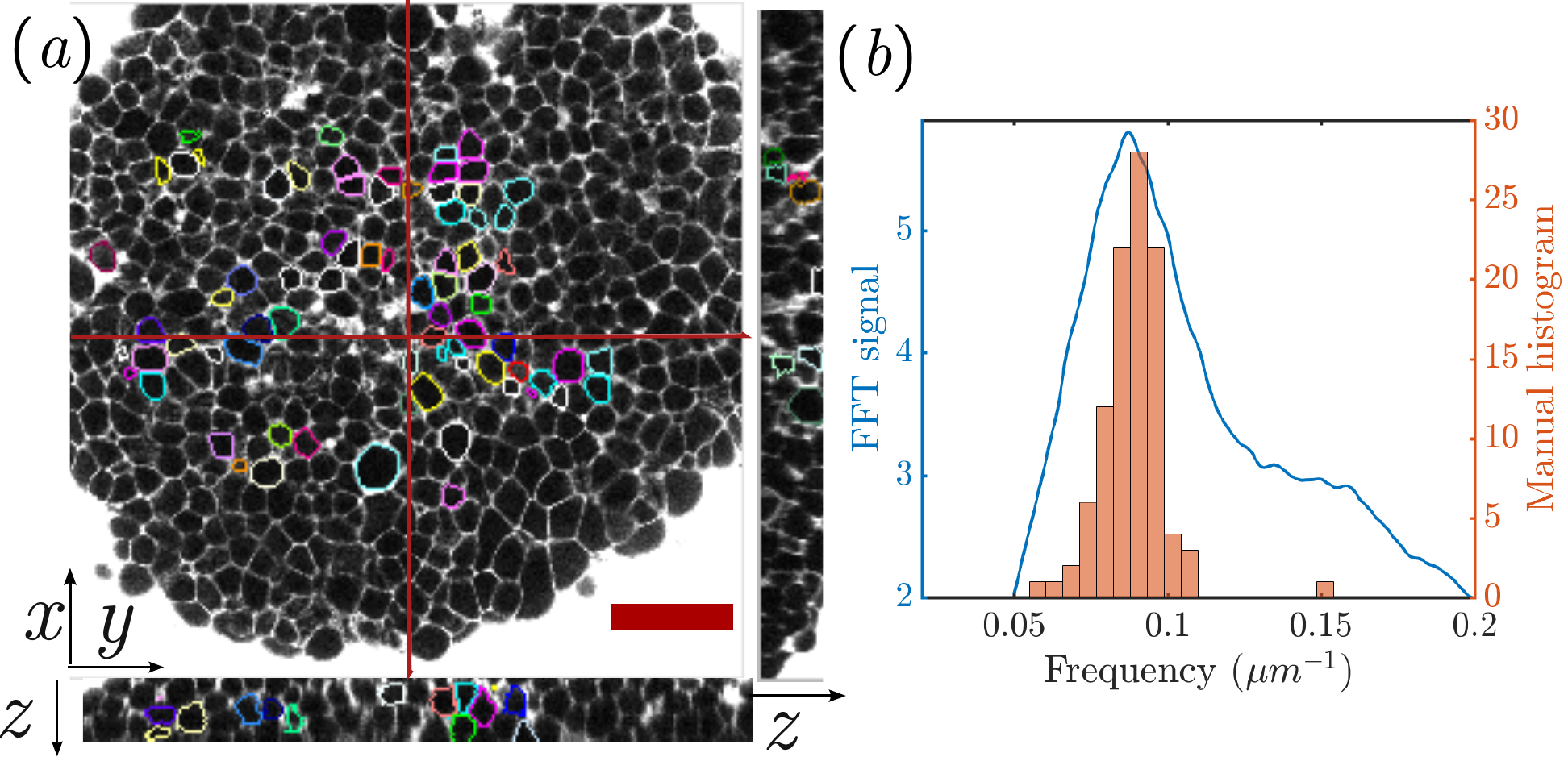}%
  \caption{$(a)$ Projections of a  two-photon stack of  a cellular aggregate in (xy), (xz), and (yz). Bar: 50 $\mu$m. Cells segmented manually using Amira software are encircled with colors. $(b)$Superimposition of the histogram of $1/D_V$ obtained manually (orange bins) and of $1/D_F$ obtained with the FFT signal (blue line). }    \label{fig:real}
\end{figure}
The performance of the algorithm was then tested on real data. Cellular aggregates of HT29 cells were exposed to  a fluorophore which only penetrates in the interstitial space. Two-photon microscopy was used for imaging (Fig. \ref{fig:real}). The  energy density function  (Eq. \ref{eq:I2agregat}) was computed (a few seconds of computation). In parallel, 100 cells of the dataset were manually segmented using Amira sofware (a few hours of segmentation)  (Fig. \ref{fig:real}a). Figure \ref{fig:real}b shows that the FFT analysis fits remarkably well with the calculation from the manual segmentation. The mean size of the manually segmented cells was of $11.40 \pm 1.38 \mu m$ (average value $\pm$ standard deviation) whereas the measurement from the Fourier analysis gave $11.50 \pm 1.37 \mu m$. The latter standard deviation was obtained following Eq. (\ref{eq:sF}) as $\sigma_F/5.4$.

\section{Sensitivity of the experimental measurement to the spherical symmetry}
\begin{figure*}
  \includegraphics[width=\textwidth]{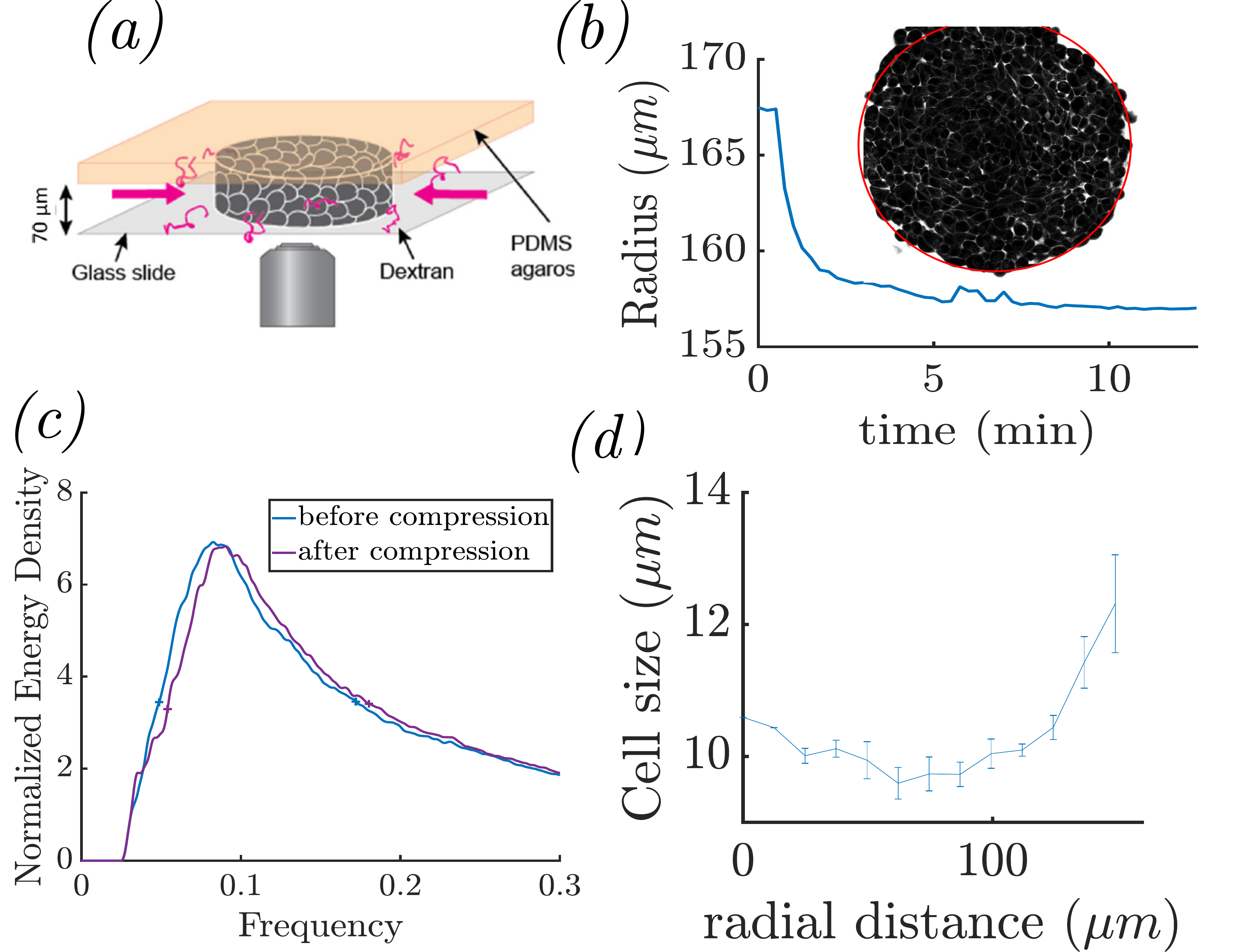}%
  \caption{$(a)$ A HT29 aggregates is confined in a microfluidic chamber and submitted to an osmotic shock as done in \cite{Dolega2021}. $(b)$ The aggregate radius is tracked during the shock. After segmentation a circle is fitted to the aggregate global shape. The ratio of the radii before and after the shock is $R_{after}/R_{before}=0.937$. $ (c)$ Normalized energy density as a function of frequency before (blue) and after (violet) the osmotic shock.  Difference in size is identical to the macroscopic measurement : $D'_{after}/D'_{before}=0.937$.}    \label{fig:osmoticshock}
\end{figure*}

The sensitivity of the FFT analysis to the spherical shape of the aggregate was tested. A cellular aggregate was positioned in a microfluidic channel. Confinement in the channel imposed a cylindrical shape (Fig. \ref{fig:osmoticshock}a). The in-plane radius was examined in response to an osmotic shock  (see \cite{Dolega2021} for details). Variation of the osmotic pressure provoked a decrease in size of the whole aggregate and of the individual cells. We tested the FFT analysis to quantify the difference in cell size  assuming that the number of cells remains constant in the aggregate during the experiment (which is a valid assumption as the compression only lasts half an hour). First the initial radius of the spheroid $R_{before}$ and its final radius after compression  $R_{after}$ were measured. A ratio of  $R_{after}/R_{before}=0.937$ was obtained (Fig. \ref{fig:osmoticshock}b). Ratio of the mean cell sizes obtained with the FFT analysis gave consistent results:  $D'_{after}/D'_{before}=0.937$ ((Fig. \ref{fig:osmoticshock}c). This result then showed that (i) the FFT analysis can handle a cylindrical symmetry, and (ii) it allows addressing variations of the size of few $\%$.

\section{Conclusions}
We present here an original method based on Fourier Transform to calculate the mean cell size and the width of its distribution very accurately. This method exploits the fact that the Fourier transformation of the fluorescent signal gives a pattern that can be analyzed with tools from elastic scattering theory. A first observation is the isotropic scattering pattern of spherical symmetry, which supports the intuition that cells in 3D aggregates have a weak positional order. The originality of the method is to analyze the power spectral density not renormalized by the spectral volume. This allows enhancing signals that are away from the center of the scattering pattern. Then the position of the first peak can be accurately measured, giving access to an accurate quantification of the cell mean diameter. While in principle the scattering pattern could provide finer information on the mean form of the cells or on their positional order, these informations rely on the use of interaction models. To our best knowledge, such models do not exist for cellular organizations yet, thus limiting the exploitation of the diffusion pattern. An empirical approach was however proposed to obtain the width of the size distribution, which will gain being confirmed by theoretical modeling. 

 The accuracy of this new tool was tested thoroughly in regards of experimental conditions. For instance, we show that the a volume of $3\times3\times 3$ cells is enough for obtaining an accurate quantification of the cell size, with error below 5\%. Thus it does not necessitate a very high accuracy in z resolution nor a very large field of view.  It is interesting to note that one single set of data can be used several time with different sectioning (varying dz for example) to get different measurements of the peak location. Then, the averaged result gives an even more accurate quantification. There is no doubt that this technique can be applied also \textit{in vivo} in situations where  automatic cell segmentation is too difficult as in the data presented here. Further development could enclose the study of cell anisotropy, which would necessitate fitting the data by an ellipse instead of a sphere.

\begin{acknowledgements}
The project was supported by the Labex Imust (Universit\'e Lyon). We thank C. Moulin from Nanoptec Center (Universit\'e Lyon) for help on two-photon data acquisition.
\end{acknowledgements}
S. M. and H. D.-A. designed the experiment. T. H., S. M. and C. J. performed the experiments. T. H., S. M. and H. D.-A. analyzed the data. A. N. and H. D.-A. interpreted the data. T. H., A. N. and H. D.-A. wrote the manuscript.

\bibliographystyle{spphys}       

\end{document}